\documentclass[final,5p,times,twocolumn]{elsarticle}
\usepackage{graphicx}
\usepackage{bm}
\usepackage{psfrag}
\usepackage{amssymb}
\usepackage{amsthm}

\begin{document}
\begin{frontmatter}
\title{Nonlinear doping of cuprate superconductors -- The case of Bi$_2$Sr$_{2-x}$La$_{x}$CuO$_{6+\delta}$}
\author{J. R\"ohler}
\ead{juergen.roehler@uni-koeln.de}
\ead[url]{http://www.uni-koeln.de/$\sim$abb12}

\address{Fachgruppe Physik, Universit\"at zu K\"oln, Z\"ulpicher Str.77,  50937 K\"oln, Germany}

\begin{abstract}
We analyze the hole doping mechanism in 
Bi$_2$Sr$_{2-x}$La$_{x}$CuO$_{6+\delta}$ (BSLCO). 
The singular optimum around $x=0.35$ is found to be connected with a 
feedback between the doped CuO$_2$ layers and its dopant  reactant 
[La$^{3+}$/Bi$^{3+}$--O$_{\delta}$] locking the number of doped holes  
preferentially on to the universal optimum $n_{opt}\simeq 0.16$. 
\end{abstract}


\end{frontmatter}

The effective charge metallizing the CuO$_2$ layers in cuprate
superconductors must be generated spatially outside in one of the 
functional layers or combinations of them, usually labelled 
``separating'', ``insulating'', and ``spacing''.
Thus the doping mechanism rendering cuprates superconducting is
fundamentally different from that in high-$T_c$ band metals, $e.g.$
the superconducting iron pnictides.  In the latter $3d$~transition
metal or anion doping is reproducibly successful, in the former it
has an deterious effect on the superconductivity.  By choosing which
combinations of outside layers are used for dopant cations and/or oxygen 
atoms, $T_c$ of the cuprates may be raised up to 164~K, three times higher
than in high-$T_c$ band superconductors.  Oxygen dopants, residing in
the strongly polarizable ``insulating'' layers far apart from
unbuckled CuO$_2$ sheets, were found to be most efficient enhancing
$T_c$.  The relationship between $T_c$ and the number of carriers in
$p$-type cuprate superconductors is universal in that $T_c$ passes a
common maximum at $n_{opt}=0.16\pm 0.01$, independent on the variety of
the chemical and structural complexes hosting superconducting
CuO$_2$ layers \cite{GroFei}.  $1 - T_{c}/T_{c}^{max}=82.6(n-0.16)^2$
came into widespread use as ``universal'' parameterization of the 
the superconducting ``dome'' of hole doped cuprates. Truly
universal, however, seems to be only the optimum hole number, 
$n_{opt}=0.16$. The ``domes'' are generally distorted by
striking dips (e.g. ubiquitously at $n=1/8$),  a cusp at the optimum, 
and by a strong asymmetry in the overdoped regime.  And the nominal edges 
at the under ($n=0.05$) and overdoped ($n=0.27$) sides 
are frequently found very close to the 
optimum, most prominently in La/Bi-doped BSLCO. 

Superconductivity of BSLCO starts only at $n\geq 0.10$ and fades away 
already at $n\simeq 0.22$.  A narrow asymmetric cusp in $T_c (n)$, 
unvariably centered at $n_{opt}=0.16\pm 0.01$ \cite{OnoAnd, SchMan}, 
is unaffected by different crystal growth routes, varying Pb$^{+3}$/Bi$^{+3}$ substitutions, 
and maxima ranging from 18 ~K in poly to 34~K in monocrystalline
samples.  It is noteworthy that optimum doping maximizes not only 
$T_c$ but also the volume fraction of the Meissner effect.  Significant changes of
the atomic structure occur at the optimum: the buckling of the CuO$_2$
layers gets minimized \cite{RoeKra} and the symmetry of the
superlattice cell uniformly orthorhombic, forcing out intergrown 
lamellae of underdoped monoclinic supercells \cite{MarDud}.  Also the sharply peaking 
low-$T$ Hall number exhibits singular behavior at $n_{opt}=0.16$ \cite{BalBoe}.  
Apparently the electronic and atomic structures of
BSLCO lock on to the superconducting optimum. Noteworthy, 
the singular optimum is robust on a high energy scale. It is hard to reconcile 
it with a quantum critical point, and the rugged $T_c (n)$ 
relationship simply with the borderline of a quantum-critical fluctuation regime.

In this brief note we summarize a collaborative experimental effort 
\cite{RoeDwe} on BSLCO, bringing out that the doping mechanism to be 
nonlinear,  not just linear charge transfer. The composition and the 
structure of the reactant doping optimally the CuO$_2$ layers, is 
itself constrained by the optimum $n_{opt}=0.16$.

La$^{3+}$/Sr$^{2+}$ substitution in BSCO is chemically correlated with
the insertion of excess oxygen atoms between the ``insulating''
Bi$^{3+}$--O(4), and ``separating'' La$^{3+}$/Sr$^{2+}$--O(3) layers.  The
excess oxygen atoms may reside at interstitial ``A'' and ``B'' sites, close to
the octahedral apex (O3) and to the octahedral faces, respectively. 
La$^{3+}$ bonds both, O$^{\mathrm A}$ and O$^{\mathrm
B}$, hence building an effective doping reactant:
[O$^{\mathrm A}$-2La-O$^{\mathrm B}$]$_{0.5}$.  BSLCO forms
in the entire superconducting regime and is almost unaffected by post-annealing, 
a favorite system for experimentalists! 
O$^{\mathrm A}$ and O$^{\mathrm B}$ have different electronic functions.  While
O$^{\mathrm A}$ may oxidize the CuO$_{2}$ units to Cu$^{\mathrm {III}}$, 
O$^{\mathrm B}$ may compensate just the La$^{3+}$/ Sr$^{2+}$ charge contrast.

\begin{figure}
\includegraphics[width=8cm,clip=true]{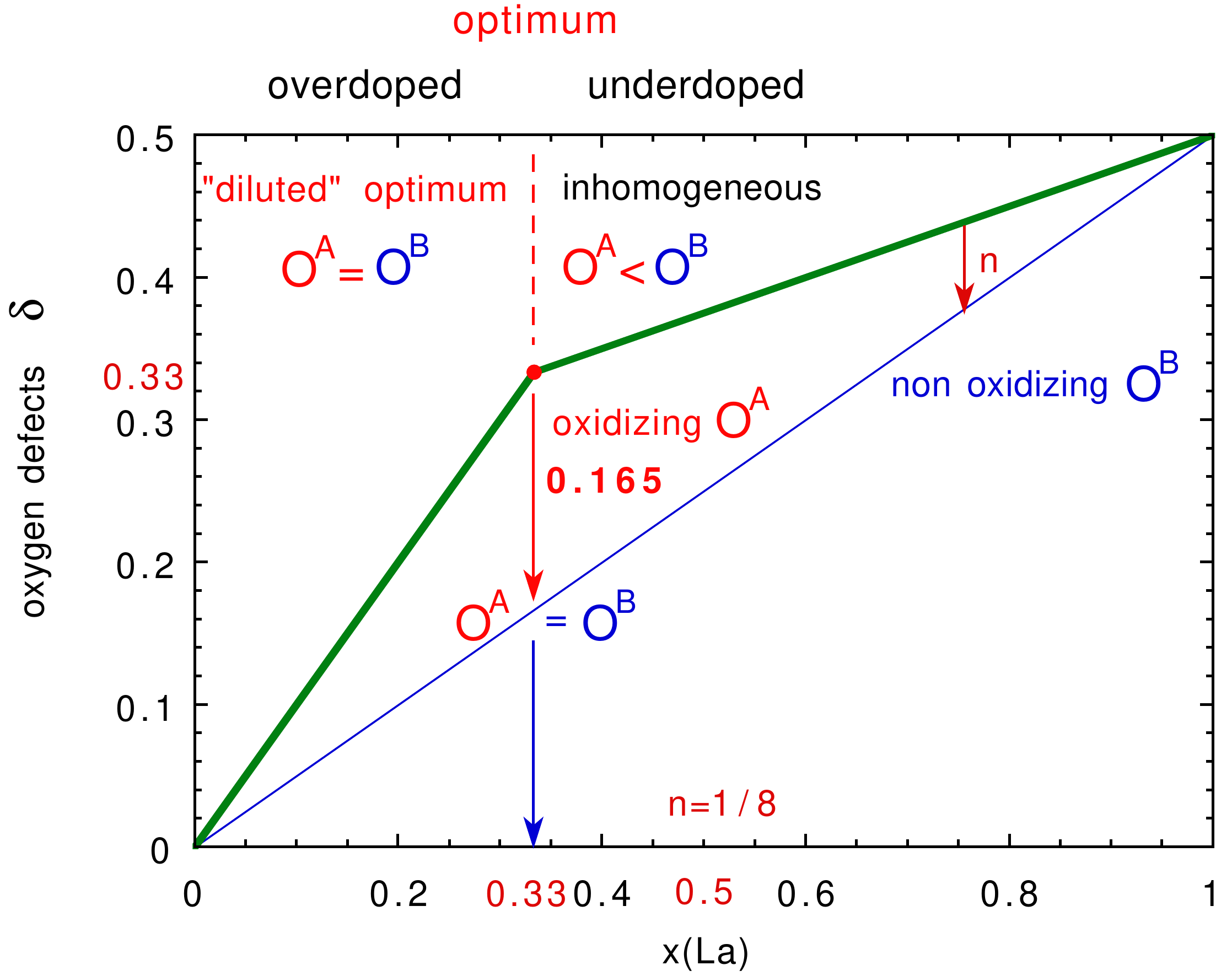}
\caption{Number of oxygen excess atoms at interstitial 
A and B sites as a function of La in Bi$_2$Sr$_{2-x}$La$_x$CuO$_{6+\delta}$. 
O$^{\mathrm B}$ compensates the La/Sr charge contrast, 
O$^{\mathrm A}$ oxidizes the CuO$_2$ layers. 
Resulting hole numbers $n$ are indicated by the vertical arrows. 
The universal optimum $n_{opt} \simeq 0.16$ occurs for $x=0.33$. 
Note the paradoxical behavior in the overdoped regime $x\le 0.33$.} 
\label{fig1}
\end{figure}

\begin{figure}[thb]
\includegraphics[width=8cm,clip=true]{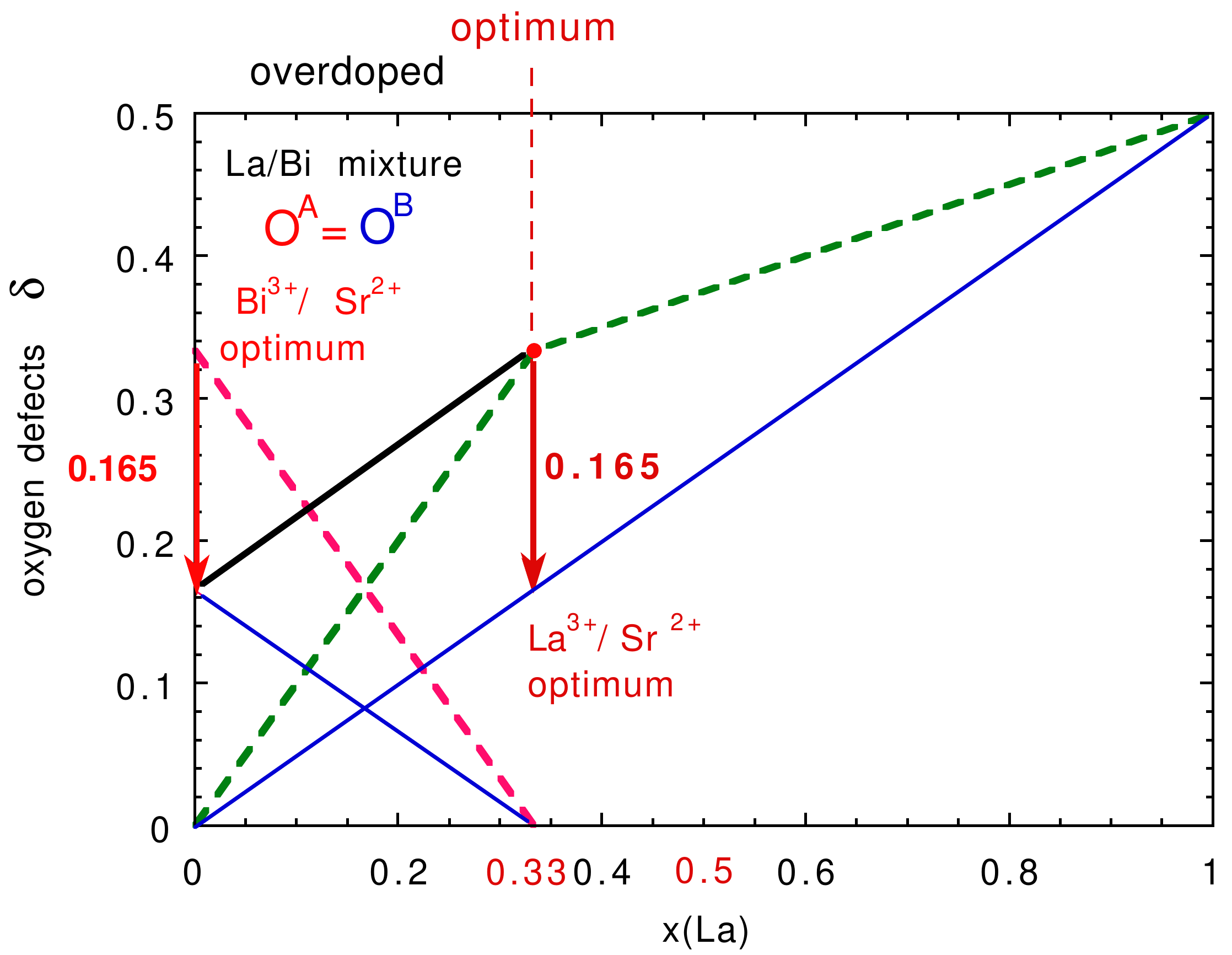}
\caption{Analysis of the ``hole saturation'' (parallel 
thick lines) observed in the overdoped regime. Optimally doped CuO$_2$ layers 
are singular and tend to repell holes from overdoping [La-O] reactants. 
A lever rule governs the ratio of mixed BSBCO and BSLCO phases.}
\label{fig2}
\end{figure}

We model in Figs.~\ref{fig1} and \ref{fig2} the relationship between the
lanthanum concentration $x$ and the total number of oxygen excess
atoms, $\delta=\delta_{\mathrm {O^A}}+\delta_{\mathrm {O^B}}$, 
by a tilted ``roof\/'' (thick green lines).  
It peaks at $\delta=0.33$ for $x=0.33$, well within the 
experimentally established range of the optimum: $0.38\ge x_{opt}\ge 0.3$.
The peak position is obtained from two intersecting straights: 
$\delta(x) =0.5(1-x)$ (not shown), and $\delta(x) =x$. 
The La-rich end member is insulating and antiferromagnetic, 
hence accommodates only $\delta =0.5$ oxygen excess atoms in 
a nonoxidizing [2La-O$^{\mathrm B}$]$_{0.5}$ reactant.  
Removal of La$^{3+}$ from the parent compound will nominally 
redistribute up to $\delta=0.5$ oxygen excess atoms 
from B to oxidizing A~sites in the reactant [O$^{\mathrm 
A}$-2La-O$^{\mathrm B}$]$_{0.5}$.
In the La-poor regime however each added La$^{3+}$ atom drags exactly one 
O$^{2-}$ forming [O$^{\mathrm A}$-2La-
O$^{\mathrm B}$]$_{0.5}$.  Thus $\delta = x$ dopes optimally. The 
nonoxidizing fraction follows the thin blue diagonal line. As indicated by the vertical 
red arrows in Fig.~\ref{fig1} and \ref{fig2}, $x_{opt}=0.33$ 
dopes $n_{opt}=0.165$ holes.

The left side of the ``roof\/'' is constructed assuming 
the nominal Bi$^{3+}$ stoichiometry,  
leading to paradoxical ``underdoping'' in the overdoped 
regime $0\leq x\le 0.33$. But here O$^{\mathrm A}$ = 
O$^{\mathrm B}$, in contrast to the underdoped regime. 
Alike the optimum with $0.165$ holes for $x=0.33$ the overdoped 
regime has $n=0.5/x$. Notably all laboratories found excess nonstoichiometry of Bi to be 
mandatory for the safe growth of overdoped crystals. 
In fact, careful examination of their actual 
stoichiometry brings out seizable excess of Bi$^{3+}$.
  
For $x\le x_{opt}$ Bi$^{3+}$ starts to substitute Sr$^{2+}$ 
thereby expelling La$^{3+}$. The stoichiometry of overdoped BSLCO must be 
hence written Bi$_{2+y}$Sr$_{2-x-y}$La$_{x}$CuO$_{6+\delta}$. 
$y=\delta_{0}-x$ may rise in La-free crystals 
up to 0.33. Dependent on the preparation  
$T_{c}^{max}$ of Bi$_2$Sr$_{2-x}$Bi$_{x}$CuO$_{6+\delta}$ (BSBCO) 
does not exceed $\simeq 10$~K. Noteworthy, fully stoichiometric
Bi$_2$Sr$_2$CuO$_6$, and even Bi$_2$Sr$_2$CuO$_{6+\delta}$ decompose. 

The special situation in the overdoped regime $0\leq x_{opt}\le 0.33$
is displayed in Fig.~\ref{fig2}. Spectroscopically (by using XAS) the hole number in the 
overdoped regime was found to ``saturate'' \cite{SchMan}, 
recently confirmed also from overdoped crystals of other cuprate families.  
Apparently the overdoped regime tends to reject all holes exceeding 
$n_{opt}\simeq 0.16$. As a consequence even the 
heavily doped system is in favor of the 
superconducting optimum. Chemical treatment attempting to force 
the system into the overdoped regime will only segregate the crystal 
into a phase mixture of optimum BSLCO with near optimum (or underdoped) 
BSBCO. The mixed system gains the mixing ratio 
as additional thermodynamical degree of freedom stabilizing it with a constant 
hole number. Single crystal X-ray diffraction has recently found strong evidence
for macroscopic decomposition in many overdoped crystals  
of BSLCO \cite{MarDud}.  Fig.~\ref{fig2} proposes a simple lever rule 
to govern the actual mixing ratio (optimum~BSLCO)~: (optimum~BSBCO) 
in the overdoped regime. Thus the total number of total number of 
oxygen excess atoms $\delta_{\mathrm {BSLCO}}+\delta_{\mathrm {BSBCO}} = 0.33$ is 
conserved in the entire overdoped regime. The hole number is fixed 
at $n_{opt}=0.165$, note the vertical arrows in Fig.~\ref{fig2}. 
Each of the two chemically different doping reactants is formed 
with equal amounts of O$^{\mathrm A}$ and O$^{\mathrm B}$. 
Due to the lower $T_{c}^{max}$ of BSBCO compared to BSLCO 
the effective $T_{c}$ of the overdoped phase mixture however decreases.

Clearly, the origin of the singular 
optimum at $n_{opt}=0.16$ has to be sought in the special correlations 
between holes doped into CuO$_2$ layers \cite{Roe}. But the doping reactant 
has to be able adapting to them by optimization of its 
effective structure through the optimally doped CuO$_2$ layers themselves.

The support by the European Synchrotron Radiation Facility 
through Projects HE2644 and HE2955 is gratefully acknowledged.


\begin{thebibliography}{00}
\bibitem{GroFei}W.A. Groen, D.M. de Leeuw, L.F. Feiner, Physica C {\bf 
165} (1990) 55.
\bibitem{OnoAnd}S. Ono, Y. Ando, Phys. Rev. B {\bf 67} (2003) 104512.
\bibitem{SchMan} M. Schneider, R.-S. Unger, R. Mitdank, R. M{\"u}ller, 
A. Krapf, S. Rogaschewski, H. Dwelk, C. Janowitz, R. Manzke, Phys. Rev. 
B {\bf 72} (2005) 014504.
\bibitem{RoeKra}J.~R{\"o}hler, M.~Klinkmann, H.~Hess, R.~Manzke, H.~Dwelk, 
A.~Krapf, Annual Report HASYLAB Deutsches Elektronen Synchrotron (Hamburg, 
Germany), Part {\bf I}, 2002, p.~507.
\bibitem{BalBoe}F. F. Balakirev, J. B. Betts, A. Migliori, S. Ono, Y. Ando, 
G. S. Boebinger, Nature {\bf 424} (2003) 912.
\bibitem{RoeDwe} J. R{\"o}hler, A. Krapf, H. Dwelk {\it et al.}, in 
press.
\bibitem{MarDud}V.P.~Martovitsky, A.~Krapf, L.~Dudy, JETP Lett. {\bf 
85} (2007) 292.
\bibitem{Roe} J. R{\"o}hler,  Physica C {\bf 460-62} (2007) 374.

\end{thebibliography}
\end{document}